\documentclass[aip,apl,twocolumn]{revtex4}
\usepackage{graphicx}
\begin{document}

%




\title{Making a field effect transistor on a single graphene nanoribbon by selective doping}

\author{Bing Huang, Qimin Yan, Gang Zhou, Jian Wu}
\affiliation{Department of Physics, Tsinghua University, Beijing
100084, China}
\author{Feng Liu}
\affiliation{Department of Materials Science and Engineering,
University of Utah, Salt Lake City, Utah 84112}
\author{Bing-Lin Gu, and Wenhui Duan\footnote{Author to whom correspondence should be addressed;
Electronic mail: dwh@phys.tsinghua.edu.cn}}
\affiliation{Department of Physics, Tsinghua University, Beijing
100084, China}
\date{\today}

\begin{abstract}
Using first-principle electronic structure calculations, we show a
metal-semiconductor transition of a metallic graphene nanoribbon
with zigzag edges induced by substitutional doping of Nitrogen or
Boron atoms at the edges. A field effect transistor consisting of
a metal-semiconductor-metal junction can then be constructed by
selective doping of the ribbon edges. The current-voltage
characteristics of such a prototype device is determined by the
first-principle quantum transport calculations.

\end{abstract}

\maketitle



Graphene nanoribbons (GNRs) have attracted intensive interest
because of their unique electronic properties and vast potential
for device applications\cite{Nakada, Wakabayashi, Kusakabe, Louie,
T. B. Martins, Qimin, Pisani, Kim, Chen, White-circuit, 0709.1163,
Fiori}. In particular, the GNR-based devices could behave like
molecule devices, such as those based on carbon nanotubes
(CNTs)\cite{Avouris, Javey, Weitz}, but with some inherent
advantages, including more straightforward fabrication processes
by using lithography technique and better control of
crystallographic orientation in constructing device
junctions\cite{Qimin,Kim,Chen,Fiori}. Different from CNTs, the
existence of edge structures endows GNR with some novel physical
and chemical properties, such as the high edge reactivity
\cite{Jiang} and unique edge states around the Fermi
level\cite{Nakada}. These may offer key advantages in realizing
various electronics applications via edge chemical
functionalization, such as {\it doping}.

It is well known that Nitrogen (N) and Boron (B)atoms are typical
substitutional dopants in carbon materials (like CNTs\cite{Jnn5}),
and their binding with the C atom is covalent and quite strong,
comparable to that of host C-C bond. The incorporation of N or B
atoms into the carbon materials will influence the electronic and
transport properties of the C host by introducing extra carries
and/or new scattering centers\cite{0705.3040}. In this letter, we
theoretically show that a metal-semiconductor transition (MST) can
be induced in an ``armchair'' GNR (with zigzag edges)\cite{Qimin}
by substitutional doping of N or B atoms on the edges.  Based on
this finding, we propose a field effect transistor (FET) made from
an single armchair GNR via selective edge functionalization
(doping), and demonstrate that the characteristics of such a FET
is comparable with that of the CNT-FETs.

Our electronic structure calculations are performed using the
Vienna {\it Ab-initio} Simulation Package \cite{Kresse}, which
implements the formalism of plane wave ultrasoft pseudopotential
based on density functional theory (DFT) within local density
approximation (LDA). The plane wave cut-off energy is set as 350
eV. Structural optimization was first carried out on all doped
systems until the residual forces on all ions were converged to
below 0.01 eV/{\AA}. The quantum transport calculations were
performed using the Atomistix ToolKit2.0 package
\cite{Taylor,Brandbyge}, which implements DFT-based real-space,
nonequilibrium Green's function formalism. The mesh cutoff of
carbon atom is chosen as 100 Ry to achieve the balance between
calculation efficiency and accuracy.

A (2, 2) armchair GNR with H termination (Fig. 1a) is chosen as a
model system to study the effect of substitutional doping of N and
B atoms in armchair GNRs. Herein, the ``armchair'' GNRs are
denoted using a nomenclature in analogy to armchair carbon
nanotubes that would unfold into corresponding ribbons with zigzag
edges\cite{Qimin}. Our calculations show that the zero-temperature
ground state of armchair GNRs is spin-polarized in agreement with
previous calculations\cite{Louie}. The energy of the
spin-polarized state is only ~20 meV per edge atom lower than the
spin-unpolarized state. However, the spin-polarized state would
become unstable with respect to the spin-unpolarized state in the
presence of a ballistic current through the
GNRs\cite{White-circuit}. Moreover, the magnetization was shown
theoretically to be forbidden in pure 1D and 2D systems at finite
temperatures \cite{Mermin}, while most transistors work at finite
temperature (room temperature). Therefore,  we will only consider
the spin-unpolarized state of armchair GNRs for our investigation
of GNR-based devices. In this case, pristine armchair (2,2) GNR is
metallic with partially flat bands at the Fermi energy (the
so-called ``edge states''\cite{Nakada}) localized at the ribbon
edges  (Fig. 1b).

\begin{figure}[tbp]
\includegraphics[width=8.5cm]{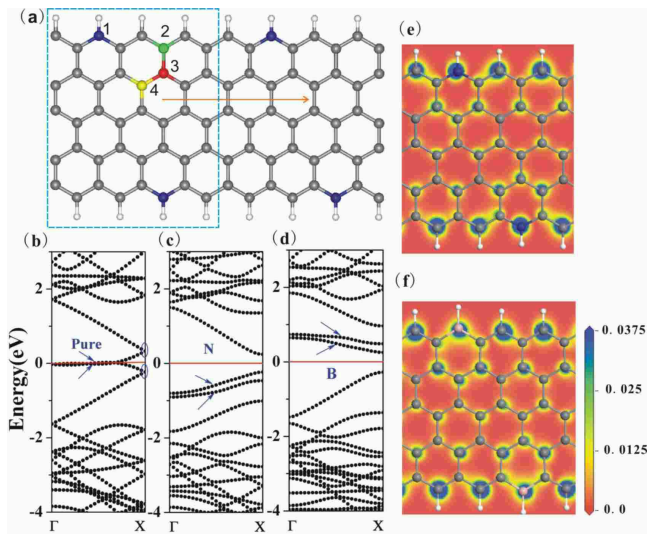}
\caption{(Color online). (a) Atomic structure of the armchair
(2,2) GNR, where the arrow shows the periodic direction. The edges
are terminated by the H atoms (denoted by small white spheres).
Four different substitutional sites are considered. (b) Band
structure of pure (2,2) GNR. The Fermi energy is set to zero and
the ``edge states'' are indicated by the arrows. (c) Band
structure of the N-doped (2,2) GNR. (d) Band structure of the
B-doped (2,2) GNR. The partial charge density of two bands
labelled by the two arrows of the N-doped and B-doped GNRs are
shown in (e) and (f), respectively. The scale bar is in units of e
\AA$^{-3}$. }
\end{figure}

Doping is achieved in the supercell made of the 4 unit cells by
substituting a N or B atom for a C atom in the GNRs. Four
different doping sites are considered as shown in Fig. 1a, and the
corresponding total energies are calculated to determine the most
energy-favorable site. For N doping, the calculated substitution
energy of site 1 is much lower than those of site 2 (by 1.07 eV),
site 3 (by 1.00 eV) and site 4 (by 1.32 eV). While for the case of
B doping, the corresponding energy differences are 0.69 eV, 0.60
eV and 0.97 eV, respectively. This clearly indicates that the edge
(site 1) of GNR is the most energetically favorable site for N or
B substitution. Furthermore, it is found that the substitution of
N or B atoms for C atoms does not affect the stability of overall
configuration, consistent with previous experimental results
\cite{Hoffman}. The local structural distortion induced by
B-doping is more pronounced than N-doping, like the case in CNTs
\cite{Jnn5}, which can be related to the atomic radius difference
between N (or B) and C atom.

Considering the fact that both edges of a GNR are identically
active for doping and it is difficult in practice to realize
selective doping merely on one edge while keeping another edge
unchanged, we will focus our study on the cases where two carbon
atoms (one per edge) in the supercell are simultaneously
substituted by two N or B atoms. The two substitutional edge sites
are determined by minimizing the total energy of the system, which
are indicated by blue atoms in Fig. 1a. Figs. 1c and 1d show the
band structures of the N-doped and B-doped (2,2) GNRs,
respectively. From the partial charge density analysis (shown in
Figs. 1e and 1f), we find that the energy states (indicated by two
arrows in Figs. 1c and 1d) near the Fermi energy are mainly
localized at two zigzag edges, and are derived from the original
edge states. Most importantly, the substitutional doping of the N
(or B) atoms have removed the degeneracy of the eigenvalues at $X$
point and thus open an energy gap. The N (B) substitution
consequently changes the original band filling [the two states
denoted by the arrows are now fully occupied (unoccupied)] and
eventually induces a transition of the armchair GNR from metallic
to semiconducting. This phenomenon is rather interesting and
somewhat unexpected since impurity doping, in general, results in
a transition of semiconducting to metallic. We have examined the
cases for different substitutional sites of two N (B) atoms in
GNRs and observed similar MST in the system. It should be noted
that our above calculations correspond to a uniform impurity
distribution since the conventional periodic boundary condition is
adopted and there is only one impurity at each edge in the unit
cell. In order to clarify the effect of random substitution, we
have also studied electronic structure of GNRs using a larger
supercell containing two impurities at each edge, where the
``nonperiodic'' substitution could be partially considered with
different configurations of the impurities. It is found that
compared with the periodic substitution mentioned above, such
``nonperiodic'' substitution is energetically unfavorable, and
importantly, does not change the MST of GNRs in essence. The above
results show that the edge doping-induced transition could be
general in metallic armchair GNRs.

\begin{figure}[tbp]
\includegraphics[width=8.5cm]{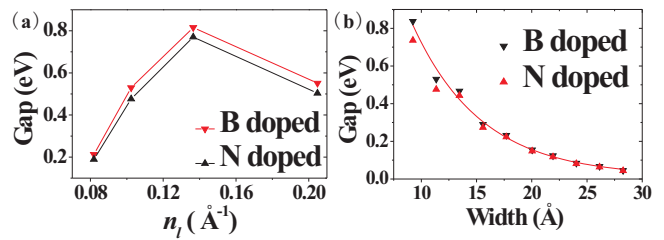}
\caption{(Color online). (a) The dependence of the band gap on the
linear doping concentration for the N- and B-doped (2,2) GNR. (b)
The dependence of the band gap on the GNR width with the linear
doping concentration of 0.1024 \AA$^{-1}$.}
\end{figure}

Next we will investigate the effect of the doping concentration
and nanoribbon width on the electronic properties of the doped
GNRs. Herein we define the linear doping concentration as: $n_l$
=$N_{\rm dopant}$/$L$, where $N_{\rm dopant}$ is the number of
dopants per supercell and $L$ is the length of the supercell along
the nanoribbon. Fig. 2a shows the band gaps of N-doped and B-doped
(2,2) GNRs as a function of the linear doping concentration $n_l$.
Note that the configuration we studied at each doping
concentration is always chosen as the energetically most favorable
one. Interestingly, it is found that the band gap first increases
with increasing $n_l$ until it reaches the maximum (0.77 eV and
0.82 eV for N-doping and B-doping, respectively) at $n_l$ of
0.1365 {\AA}$^{-1}$, and then decreases with further increasing
$n_l$. The GNRs with B substitution have slightly larger band gaps
than those with N substitution. In addition, the band gap of the
GNR decreases with increasing ribbon width and eventually
diminishes when the width is too big(Fig. 2b).

Among many challenges for the use of graphene or nanoribbon in
FETs, an important and practical issue is to fabricate
semiconducting channel with large enough band gap, which is
crucial for effectively reducing the leakage current and improving
the critical performance parameters such as ON/OFF current ratio.
However, until recently it is still very hard to experimentally
fabricate semiconducting GNRs with the energy gap larger than 0.2
eV \cite{Kim,Chen}. The doping-induced MST we report here may be
used to provide another way to fabricate transistor semiconducting
channels in future graphene based functional devices. Below we
will demonstrate the characteristics and performance parameters of
the N-doped GNR-FET by first-principle quantum transport
calculations. It should be noted that both the electrodes and the
conduction channel are integrated on a single armchair GNR in such
a device (Fig. 3a). Such a linear configuration can also be
advantageous to increase the device density in an electronic
circuit as well as to simplify the fabrication process.

\begin{figure}[tbp]
\includegraphics[width=8.5cm]{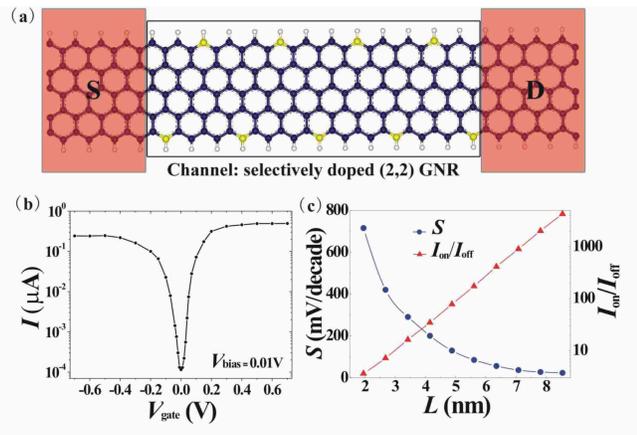}
\caption{(Color online). (a) The schematic structure of the field
effect transistor (FET) made from a single (2,2) GNR. The
semiconducting channel is obtained by edge doping of N in a
finite-length region (the center region). (b) Simulated $I-V_{\rm
gate}$ curves of N-doped GNR-FETs under $V_{\rm bias}$ = 0.01 V.
The channel length is 8.54 nm and the linear doping concentration
is 0.1365 \AA$^{-1}$. (c) The dependence of the subthreshold swing
$S$ (blue line) and the ON/OFF current ratio (red line) on the
channel length $L$.}
\end{figure}

Fig. 3b shows the typical $I$-$V_{\rm gate}$ curves for the
N-doped GNR-FET (with the channel length of 8.54 nm) at the bias
voltage $V_{\rm bias}$=0.01 V. In the voltage window -0.7
V$\sim$0.7 V, the doped FET exhibits ambipolar characteristics
with the on-current ($I_{on}$) of $\sim$ 1 $\mu$A. The minimum
leakage current is limited to a rather small value ($\sim$
1.2$\times$10$^{-4}$ $\mu$A), and a high ON/OFF current ratio
($I_{\rm on}/I_{\rm off}>$ 2000) is achieved in such N-doped
GNR-FETs. The large ON/OFF current ratio manifests the ``perfect''
atomic interface between the metal-semiconductor GNR junctions
with a minimum contact resistance. It increases the possibility of
experimental operation between ON and OFF states. Moreover, such a
device exhibits an excellent ``theoretical'' switching
characteristics with a subthreshold swing $S=\ln(10)[dV_{\rm
gate}/d(\ln I)]$ $\sim$ 40 mV/decade. This is comparable to that
of high performance CNT-FET (60-80 mV/decade) \cite{Javey,Weitz}
and reaches the theoretical limit of $S$ ($\sim$ 60 mV/decade) for
Si-based FET at room temperature \cite{Sze}. The switching
mechanism here can be understood in terms of a semiclassical band
bending mechanism \cite{Qimin}.

We further studied the relationship between the device performance
and the channel length by calculating $I$-$V_{\rm gate}$ curve of
N-doped GNR-FETs as a function of the doped channel length from
0.49 nm to 8.54 nm while keeping the bias voltage $V_{\rm bias}$
at 0.01 V. As shown in Fig. 3c, the subthreshold swing $S$ of
these doped GNR-FETs decreases and the ON/OFF current ratio
increases exponentially. Our calculations show that in order to
obtain good device performance with small $S$ value (e. g., below
100 mV/decade) and high ON/OFF current ratio (e. g., above 100),
the doped channel length needs to be longer than 5 nm. The minimum
leakage current of those FETs with the doped channels shorter than
this critical length will be greatly enhanced by direct tunneling,
which lowers the device performance.

In conclusion, using first-principle calculations, we have studied
the electronic and transport properties of armchair GNRs with
substitutional edge doping of N or B atoms. It is found that the
edge doping will greatly modify the band structure (especially the
``edge states'') of the system, and induce a metal-semiconductor
transition. The band gap of the doped GNR exhibits a strong
dependence on both the linear doping concentration and the
nanoribbon width. It is demonstrated that electronic devices, such
as FETs, could be integrated on a single GNR by selective edge
doping/chemical functionalization, which avoids the need to
connect/integrate the GNRs with different orientations. Simulated
$I$-$V_{\rm gate}$ curves indicate that such FETs exhibit
ambipolar ON-OFF characteristics with excellent performance
parameters.

This work was supported by the Ministry of Science and Technology
of China (Grant Nos. 2006CB605105 and 2006CB0L0601), and the
National Natural Science Foundation of China (Grant Nos. 10325415,
10674077 and 10774084). One of the authors (F. Liu) acknowledges
support from DOE.


\begin{references}

\bibitem{Nakada} K. Nakada, M. Fujita, G. Dresselhaus, and M. S.
Dresselhaus, Phys. Rev. B. {\bf 54}, 17954 (1996).

\bibitem{Wakabayashi} K. Wakabayashi,  M. Fujita,  H. Ajiki,  and  M. Sigrist, Phys. Rev. B. {\bf 59}, 8271 (1999).

\bibitem{Kusakabe} K. Kusakabe and M. Maruyama, Phys. Rev. B. {\bf 67}, 092406 (2003).

\bibitem{Louie} Y. -W. Son, M. L. Cohen, and S. G. Louie, Phys. Rev. Lett. {\bf 97}, 216803 (2006).

\bibitem{T. B. Martins} T. B. Martins, R. H. Miwa, Antonio J. R. da Silva, and A. Fazzio, Phys. Rev. Lett. {\bf 98}, 196803 (2007).

\bibitem{Qimin} Q. M. Yan, B. Huang, J. Yu, F. W. Zheng, J. Zang, J.
Wu, B. L. Gu, F. Liu and W. H. Duan, Nano Lett. {\bf 7}, 1469
(2007)

\bibitem{Pisani} L. Pisani, J. A. Chan, B. Montanari, and N. M. Harrison, Phys. Rev. B. {\bf 75}, 064418 (2007).

\bibitem{Kim} M. Y. Han, B. Oezyilmaz, Y. Zhang, and P. Kim, Phys. Rev. Lett. {\bf 98}, 206805 (2007).

\bibitem{Chen} Z. Chen, Y. M. Lin, M. J. Rooks, and Ph. Avouris, Physica E {\bf 40}, 228
(2007).

\bibitem{White-circuit} D. A. Areshkin, and C. T. White, Nano Lett., in press (2007).

\bibitem{0709.1163} A. H. Castro Neto, F. Guinea, N. M. R. Peres, K. S. Novoselov, and A. K.
Geim, cond-mat/07091163.

\bibitem{Fiori} G. Fiori and G. Iannaccone, IEEE Electron Device Lett. {\bf 28}, 760 (2007).

\bibitem{Avouris} Ph. Avouris, Acc. Chem. Res. {\bf 35}, 1026 (2002).

\bibitem{Javey} A. Javey, J. Guo, D. B. Farmer, Q. Wang, D. Wang, R.
G. Gprdon, M. Lundstrom, and H. Dai, Nano Lett. {\bf 4}, 447 (2004).

\bibitem{Weitz} R. T. Weitz, U. Zschieschang, F. Effenberger, H.
Klauk, M. Burghard, and K. Kern, Nano Lett. {\bf 7}, 22 (2007).

\bibitem{Jiang} D. Jiang, B. G. Sumpter, and S. J. Dai, J. Phys.
Chem. B {\bf 110}, 23628 (2006).

\bibitem{Jnn5} G. Zhou and W. H. Duan, J. Nanosci. Nanotech. {\bf 5}, 1421 (2005).

\bibitem{0705.3040} N. M. R. Peres, F. D. Klironomos, S.-W. Tsai, J.
R. Santos, J. M. B. Lopes dos Santos, and A. H. Castro Neto,
Europhys. Lett. {\bf 80}, 67007 (2007).

\bibitem{Kresse} G. Kresse and J. Furthm\"uller, Comput. Mater. Sci. {\bf 6}, 15
(1996).

\bibitem{Taylor} J. Taylor, H. Guo, and J. Wang, Phys. Rev.
B {\bf 63}, 245407 (2001).

\bibitem{Brandbyge} M. Brandbyge, J. L. Mozos, P. Ordej\'{o}n, J.
Taylor, and K. Stokbro, Phys. Rev. B {\bf 65}, 165401 (2002).

\bibitem{Mermin} N. D. Mermin and H. Wagner, Phys. Rev. Lett. {\bf 17},
1133 (1966).

\bibitem{Hoffman} A. Hoffman, I. Gouzman, and R. Brener, Appl. Phys. Lett. {\bf 64},
845 (1994).

\bibitem{Sze} S. M. Sze, Physics of Semiconductor Devices (Wiley, New York,
1981).

\end{references}
\end{document}